# Infrared absorbing electron in ice-I$_h$ is trapped by a water vacancy [*]

Ilya A. Shkrob [**]

*Chemistry Division , Argonne National Laboratory,  Argonne, IL 60439*

## Abstract

Ionization of polar liquids and glasses often yields metastable electron centers collectively known as "weakly bound" (wb) electrons that absorb to the red of the ground state of the solvated/trapped electron. Formation of these species is thought to be the initial stage of electron localization in such media. Although these commonly occurring species have distinctive absorption spectra, no insight into their structure currently exists. In this Letter, we address the structure of the wb electron in low-temperature ice-I$_h$ theoretically, using a combination of mixed quantum-classical (MQC) floating set of Gaussian orbitals (FSGO) – Metropolis Monte Carlo (MC) method and "embedded cluster" density functional theory (DFT) post treatment. The comparison with the observed properties of the wb electron in ice suggests that this species is an s-like wavefunction filling a water vacancy.

PACS numbers: 78.20.Bh, 76.30.-v, 76.30.Mi

Excess ("solvated") electrons in liquids (e.g., [1-8]) and large gas phase cluster anions (e.g., [9-13]) composed of hydroxylic molecules, such as water and alcohols, are currently the subject of many experimental and theoretical studies. The resurgence of interest to such species is driven by the emergence of new experimental techniques, including ultrafast pump-probe, [14,15] time-resolved resonance Raman, [7,8] and angle-resolved photoelectron [12,13] spectroscopies as well as advances in computer modeling of their structure and dynamics (e.g., [2-6]). The ground state of a strongly bound (sb) electron in liquid water is a species in which an s-like electron wavefunction fills a solvation cavity formed by 4-8 dangling (non-hydrogen-bonding) OH groups, [1-8] with ca.

80% of the spin density residing inside the cavity and the rest shared by frontier orbitals of oxygen atoms in the solvating OH groups. [2] Apart from this ground state species, short-lived intermediate states of electron solvation/trapping, known collectively as "weakly bound" (wb) electrons [16-19] that absorb to the red of the sb species have been observed in many polar liquids [16,17,18] and glasses, [19] but the structure of this commonly occurring excess electron species has not yet been elucidated. It has been suggested [18,19] that the wb electrons are formed via localization in cavities formed by fewer OH groups and/or methyl/methylene groups of the solvent, but no direct support for such a scenario presently exists. In this Letter, we seek to determine the structure of this species in the simplest molecular system that yields such wb electrons: proton disordered, low-temperature hexagonal water ice ($I_h$).

Ice-$I_h$ is convenient for such studies due to the regularity of its structure and exceptional stability of the wb electron at low temperature. As the ice lattice exhibits large channels along the *c*-axis and spacious 12-molecule cages capable of accommodating an extra water molecule, [20] electron solvation in ice is intermediate between the solvation in liquid water and electron localization at the surfaces of gas phase water clusters. One may expect that the wb electron would occupy these preexisting cavities and bind to fortuitously oriented water molecules at their surface. Below 73 K, radiolysis of ice-$I_h$ yields two kinds of electron centers. [21] The strongly bound electron is a short-lived species that absorbs in the visible, with the absorption maximum at 1.97 eV [21] (vs. 1.7 eV for liquid water). [1] The weakly bound electron absorbs in the near infrared (with the absorption maximum at ca. 0.42 eV) and slowly decays by tunneling to a nearby hole center (such as $O^-$, OH, and $H_3O^+$) on the millisecond time scale. [21] At higher temperature (120 K), Kawabata *et al.* [22] observed a different wb species with the absorption maximum at 1.24 eV. The lifetime of the sb electron at 6 K (77 K) is 8 ns (75 ns) in $H_2O$ ice and 120 ns (5 μs) in $D_2O$ ice, [21] indicating large isotope effect on the recombination of the sb electron with the parent hole center (the hydronium ion). These short lifetimes suggest that the sb electron is localized near the site of ionization, possibly due to the lattice distortions introduced by the species generated via disintegration of the holes or local lattice melting by the heat of ionization. By contrast, the wb electron is trapped in the bulk, apparently by an abundant defect in ice. [21] Electron paramagnetic resonance (EPR) spectrum of the wb electron in $D_2O$ ice at 4.2 K has been reported by Johnson and Moulton [23] and Hase and Kawabata. [24] In $H_2O$ ice, the peak-to-peak width of the structureless, single EPR line would correspond to 8.3 G which gives the second moment of $M_2\left({}^1H\right) \approx$ 17.5 G$^2$ vs. 21-23 G$^2$ for sb electron in amorphous alkaline ices. [2]

Whereas there are many mixed quantum classical (MQC) one-electron (e.g., [4-6]) and, more recently, multi-electron [25,26] molecular dynamics (MD) calculations for hydrated electrons in liquid water, there have been no such calculations for water ice; only crude semi-continuum and dipole relaxation models have been used. Kevan [27] and Julienne and Gary, [28] following the suggestions of Buxton *et al.* [29] and Nilsson, [30] speculated that the wb electron localizes inside the ice cage as a polaron, by reorientation of the water molecules so that their dipoles point to the common center. De Haas *et al.* [21] argued that the trapping site is likely to be a VD defect (positively vested vacancy). The latter is a combination of a water vacancy (V) center and a Bjerrum (proton ordering) D-defect [20] (-OH HO-); in this VD defect, there are three dangling OH bonds.

We address the structure of the wb electron in ice using the combination of two methods: an MQC one-electron model based upon the "floating set of Gaussian orbitals" (FSGO) approach [5] that was implemented as a Metropolis Monte-Carlo (MMC) code, as this implementation is more suitable for calculations at low temperature. Electrons trapped by cages (C), vacancies (V), interstitials (I), Bjerrum D- and L-defects, [20] and positively (VD) and negatively (VL) vested vacancies were systematically examined. The MMC ensembles were subsequently used as input for density functional theory (DFT) calculations using the B3LYP/6-311++G** method in which 20-30 water molecules constituting the trapping site were "embedded" in a larger matrix of fractional point charges representing the ice lattice (corresponding to SPC water model). [32] Averaging over an ensemble of 200 configurations taken every $10^{3-4}$ cycles of the MMC calculation yielded statistically average properties of the electron centers for a given temperature. The latter was chosen to be 200 K for VD defects and 50 K for other lattice defects.

A spherical ice-I$_h$ nanocluster of 440 water molecules with the frame origin centered on the middle of an ice cage was constructed using a 13.5 Å x 15.6 Å x 14.6 Å hypercell of 96 water molecules with near-zero overall dipole moment as constructed by Buch *et al.* [33] This cluster included all water molecules within 15 Å of the frame origin. The V defects were created by elimination of one of the water molecules forming the central cage, and the I defects were created by placing a water molecule at the center of this cage. Bjerrum D- and L- defects [20] were created by changing the orientation of one of the water molecules forming the central cage and then propagating hydrogen bonds towards the nanocluster surface, in a way similar to the method described by Rick and Haymet. [34] An SPC/Fw model [32] was used for water and the electron was represented by the wavefunction $\phi$ expanded on a set of primitive Gaussian orbitals with a radial function of the type $\chi(r) \propto r^\nu \exp[-f\alpha r^2]$ arranged in $s$-, $p$-, and $d$- shells centered on $X$, the center of mass of the electron. This center was adjusted iteratively, for every MMC step. The interaction Hamiltonian is given by pairwise additive screened Coulomb potential $Q\exp(-\lambda r^2)/r$, where $Q$ is the SPC charge on a given nucleus ($Q$=-0.82 and $\lambda$=1.0 Å$^{-2}$ and $Q$=+0.41 and $\lambda$=1.02 Å$^{-2}$ for H) and $\lambda$ =1.0 and 1.02 Å$^{-2}$ for O and H, respectively. [5] This form of the pseudopotential makes the evaluation of one-center integrals especially efficient, [5] and the results for liquid water are in a reasonable agreement with the experiment. [5,6] Preorthonormalized, hybridized $s/p/d$ shells were constructed using Boys method and the integrals were calculated using the standard Gaussian quadratures. [35] The basis set consisted of four $s/p$, $s/p/d$, $s/p/d$, and $s$- shells with $\alpha$ =0.018 Å$^{-2}$ and $f$ =20, 10, 2, and 1, respectively. The ground state eigenfunction was used to calculate the gyration tensor, $\mathbf{G}_{ik} = \langle \phi | \mathbf{x}_i \mathbf{x}_k | \phi \rangle$ (where $x_{i,k}$ are Cartesian coordinates in a frame with origin at $X$); [2] the radius $r_g$ of gyration was defined from $r_g^2 = tr(\mathbf{G})$. This radius characterizes the spatial extent of the electron wavefunction.

To explore a larger configuration space, only water atoms within a sphere of radius 5.5 Å from the frame origin were moved. The OH bonds at the surface of this inner cluster were fixed and small random (for molecules at the interface, constrained) rotations and vibrations of *all* water molecules in this inner cluster constituted a single step of the MMC calculation. In the typical calculation, the defect was first equilibrated over $10^5$ cycles, and then two FSGO MMC calculations of the total length of $10^{4-5}$ cycles were carried out. In the first calculation, electron energy was excluded from the total energy, so the corresponding FSGO solution represented the excess electron in an unrelaxed structural defect (indicated in the following by the subscript "n"). In the second calculation, this term was included and the ensemble of relaxed, electron-filled defects was obtained. For comparison, we also calculated the properties of the electron in liquid water, using the MQC MD ensemble generated in ref. 2 as input. The typical orbitals for the (s-like) singly occupied and the (p-like) lowest unoccupied states are shown in the Supplement (Figures 1S-4S). [36] For VD⁻ center (in which there are three dangling OH bonds) and V⁻ center (in which there are two dangling OH bonds) the wavefunction resembles the s-electron in liquid water, as the excess electron density is confined to a small cavity formed by these dangling bonds. By contrast, the "polaron" (C⁻) wavefunction completely occupies two neighboring ice cages. The calculated orbital energies and $s \leftarrow p$ transition energies are given in Table 1 (for FSGO calculation) and Table 2 (for DFT calculation; more complete tables are given in the Supplement). At the FSGO level, only VD⁻ and VD$_n$⁻ defects have positive vertical detachment energy (VDE) of ≈0.7 and ≈0.1 eV, respectively, which are considerably lower than the value of 2.7 eV calculated for liquid water (vs. the experimental estimate of 3.3 eV given by Coe et al. [13]). The calculated absorption spectra shown in Figure 1 are broad lines composed of the three merged s-p transitions involving the three nondegenerate p-like states. For liquid water, the calculated absorption maximum is at 1.9 eV, which is ca. 0.2 eV greater than the experimental value of 1.7 eV. [1] For VD⁻ center at 200 K, the band maximum is at 1.2 eV. If the absorption spectrum of the wb electron in low-temperature ice-I$_h$ were from these s-p transitions, that would eliminate the VD⁻ center as a candidate. On the other hand, these energetics would be consistent with the 1.24 eV absorbing species observed by Kawabata at 120 K. [22] By contrast, VD$_n$⁻, V⁻, I⁻, and C⁻ centers all absorb below 1 eV, with the V⁻ center absorbing at 0.8 eV. Given that MQC calculations in general yield absorption spectra that are blue-shifted by 0.2-0.3 eV, [2-6] these defects could be candidates for the wb electron. However, only VD$_n$⁻ and V⁻ defects support the sufficient bandwidth to account for the observed extent of the experimental spectrum, with the onset of the band at 1.3 eV. [21] The position of the absorption maximum correlates with the spatial extent of the s-like wave function, as $r_g$ increases from 3.37 Å for VD⁻ to 5.8 Å for C⁻ (Table 1).

In the "embedded cluster" DFT calculations, only part of the electron density resides inside the solvation cavity (Figure 2). A fraction of the unpaired electron density spreads over the *O 2p* orbitals of the water molecules that form the cavity. This fraction ($\phi(O_{2p})$) may be quantified as explained in ref. 2; it systematically decreases from VD⁻ to V⁻ and VD$_n$⁻ to V$_n$⁻, I⁻, and C⁻ centers, in parallel with the increase in the spatial extent of the electron wavefunction (Table 2). As the protons in the water molecules acquire their isotropic hyperfine coupling constants mainly via spin bond polarization involving the unpaired electron in these *O 2p* orbitals, [2] the increased spreading of the electron wavefunction reduces these hyperfine coupling constants and, correspondingly, decreases the second moment $M_2$ of the EPR spectrum. This effect provides tight limits on the acceptable structure for the electron trapping site. In particular, the VD⁻ center is eliminated by the unrealistically large $M_2$ of 30 G² whereas the extended $V_n^-$, C⁻, and I⁻ centers yield too low estimates for this parameter (Table 2). The best match for the observed second moment is provided by the V⁻ and $VD_n^-$ centers. The VDE for the VD⁻ and $VD_n^-$ centers calculated using the DFT method is 1.55 and 0.9 eV, respectively (vs. 3.26 eV calculated for liquid water, which is in good agreement with the estimate of Coe et al. [13]), i.e. the VD defect is a deep electron trap in ice. The V⁻ defect has much lower VDE of 0.4 eV, whereas electron trapping by an unperturbed V defect is thermoneutral. Interestingly, the estimate of 1.55 eV is close to extrapolated ($n \rightarrow \infty$) VDE for large ($n$=50-200) gas-phase isomer-II $\left( H_2O \right)_n^-$ clusters observed by Neumark and co-workers, [9,10] whereas 0.4 eV is close to the VDE extrapolated for isomer-III clusters. [9] Presently, these isomers are believed to trap the electrons at their surface; however, our calculations point to the possibility that the electron is trapped *internally*, by a defect site in an ice-like cluster. As the progenitor neutral custers are generated by expansion of water into vacuum, the state and the temperature of these clusters are not known well [9] and the formation of ice-like structures is possible.

For other lattice defects in ice-$I_h$, VDE is positive, and such centers would spontaneously eject electrons back into the ice bulk. The energy of the quasifree electron in crystalline ice-$I_h$ is not known; the best estimate for polycrystalline ice at 77 K is -0.36 eV, [37] which suggests the possibility of $s \leftarrow CB$ transition in the infrared. That would be in agreement with experiment, [24] indicating that photoexcitation of the wb electron in low-temperature ice results in its instantaneous decay via recombination. The energies of s-p bound-to-bound transitions for the $VD_n^-$ and the V⁻ centers are 1.2-1.3 eV, which considerably exceeds the value of 0.42 eV estimated for the maximum of the absorption band of the wb electron. [21] As the orbital energies and the spatial extent of the p-like states do not change substantially for all of these electron centers (see Table 2 and Figure 10S in the Supplement), it is likely that the absorption band of the wb electron originates entirely from these $s \leftarrow CB$ transitions, as is the case for the ground state electron in frozen alkane glasses that exhibit very similar absorption spectra. [38]

To summarize, the energetics and the magnetic resonance data exclude the previously suggested VD$^-$ and polaron models of the wb electron in low-temperature ice, pointing instead to the V$^-$ center. While the unrelaxed VD$_n^-$ center is not excluded by these models, it seems unrealistic that relaxation of this electron center, which requires only slight movement of the three OH dangling bonds, would take tens of milliseconds. Another consideration is the availability of the parent point defect: while there is much evidence for the existence of water vacancies in low temperature ice from positron annihilation spectroscopy, [39] the VD defect is still a hypothetical species lacking experimental demonstration. [20] Thus, we suggest that the "weakly bound electron" is a (quasifree) [31] electron trapped by vacancy centers in the bulk. Such vacancies are frozen into the ice lattice as the solid is cooled below 200 K. An intriguing possibility is that the "wb electrons" observed in the early stages of electron trapping/solvation in polar glasses (on the microsecond time scale) [16-18] and liquids (on the picosecond time scale) [19] are also such vacancy-trapped electrons. Rapid molecular dynamics occurring in these media quickly eliminates these weakly bound states, whereas in low-temperature ice such centers persist over a long period of time.

The author thanks J. M. Warman, J. Cowin, and D. M. Bartels for stimulating discussions. This work was supported by the Office of Science, Division of Chemical Sciences, US-DOE under contract No. DE-AC-02-06CH11357.

**References.**

* This work was supported by the Office of Science, Division of Chemical Sciences, US-DOE under contract No. DE-AC-02-06CH11357.

** *Electronic address:* shkrob@anl.gov.

**Figure captions.**

**Figure 1**

Absorption spectra (that is, histograms of oscillator strengths $f_{osc}$) for (i) liquid water at 300 K, (ii) VD$^-$ and (iii) VD$_n^-$ centers in ice-I$_h$ at 200 K and (iv) V$^-$, (v) Γ, and (vi) C$^-$ centers in ice at 50 K (FSGO MMC calculation). Dash-dot lines in trace (ii) show the three merged s-p sub-bands.

**Figure 2**

The typical isodensity map ($\pm$0.03 a.u.$^{-3}$ isosurface) for relaxed, ground-state s-like electron trapped by VD$^-$ and V$^-$ defects. B3LYP/6-311++G** calculation for a snapshot obtained using FSGO MMC method. The positive density is in light shade, the negative density in O 2p orbitals is in dark shade. The protons in dangling OH groups are indicated by open circles. More detailed orbital maps for these and other electron centers are given in the Supplement.

**Table 1.**

**Ensemble average parameters for trapped electron centers from FSGO MMC calculations.**

| Parameter, temperature | $VD^-$, 200 K | $VD_n^-$, 200 K | $V^-$, 50 K | $V_n^-$, 50 K | $\Gamma^-$, 50 K | $C^-$, 50 K |
|---|---|---|---|---|---|---|
| VDE, eV [a] | 0.7 | 0.09 | -0.3 | -0.7 | -0.61 | -0.74 |
| HO, eV [b] | -0.45 | 0.16 | 0.56 | 0.93 | 0.86 | 0.98 |
| $E_1$, eV [c] | 1.12 | 0.67 | 0.65 | 0.37 | 0.46 | 0.41 |
| $E_2$, eV | 1.21 | 0.77 | 0.76 | 0.49 | 0.53 | 0.43 |
| $E_3$, eV | 1.29 | 0.87 | 0.81 | 0.58 | 0.57 | 0.47 |
| $r_g$ (s), Å [d] | 3.37 | 4.27 | 4.39 | 5.63 | 5.35 | 5.8 |

(a) vertical detachment energy; (b) highest occupied orbital; for MQC MD water, VDE=2.7±0.4 eV, (c) transition energies to the three lowest unoccupied (p-like) states; (d) radius of gyration. A more complete table is given in the Supplement.

**Table 2.**
**Ensemble average parameters obtained from "embedded cluster" B3LYP/6-311++G** calculations using the FSGO MMC ensemble as input.**

| Parameter, temperature | VD⁻, 200 K | VDn⁻, 200 K | V⁻, 50 K | Vn⁻, 50 K | I⁻, 50 K | C⁻, 50 K |
|---|---|---|---|---|---|---|
| VDE, eV [a] | 1.55 ±0.1 | 0.9 ±0.1 | 0.41 ±0.05 | -0.04 ±0.03 | - | - |
| $M_2(^1H)$, G [b] | 31 ±8 | 20 ±6 | 20.5 ±3.5 | 12 ±3 | 7.4 ±0.4 | 5.2 ±0.5 |
| $\phi(O_{2p})$, x10² [c] | 14.6 | 8.2 | 8.0 | 3.0 | 4.4 | 3.6 |
| $r_g$, Å [d] | 3.02 | 3.85 | 3.77 | 4.47 | 4.76 | 5.05 |
| HO, eV [e] | -0.06 | 0.38 | 0.88 | 1.09 | 1.10 | 1.13 |
| $E_1$, eV [f] | 1.68 | 1.22 | 1.20 | 0.94 | 0.73 | 0.68 |
| $E_2$, eV | 1.82 | 1.37 | 1.23 | 0.97 | 1.06 | 0.99 |
| $E_3$, eV | 1.92 | 1.48 | 1.47 | 1.21 | 1.18 | 1.15 |

a) for MQC MD water ensemble, [2] VDE=3.26±0.45 eV, (b) estimated second moment of EPR spectrum (for protons), see ref. 2 for details of the calculation; (c) the fraction of spin density in *O 2p* orbitals of water molecules; (d) radius of gyration for the s-like state; (e) Kohn-Sham orbital energy for the highest occupied orbital; (f) transition energies for three s-p transitions. A more complete table is given in the Supplement.

Figure 1.

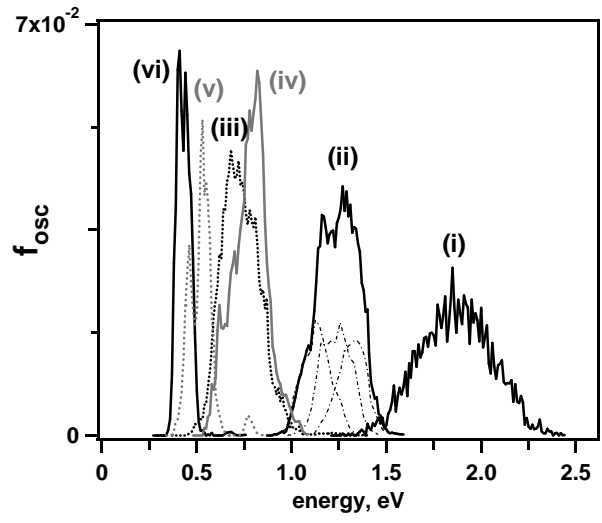

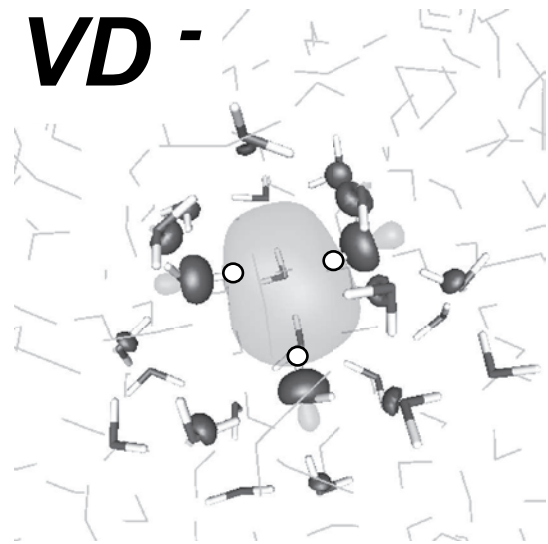

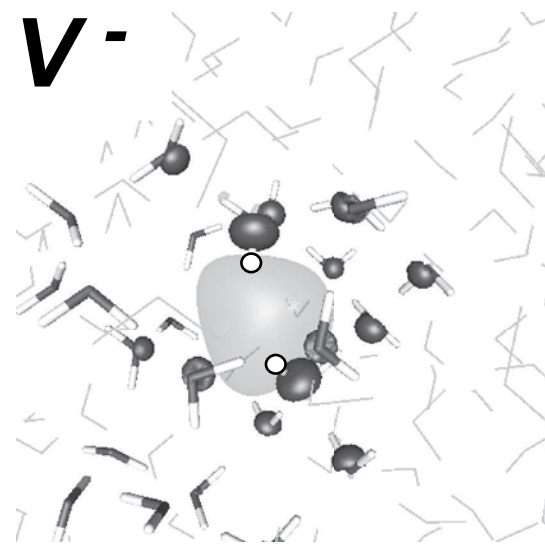

## VD⁻, 200 K : FSGO

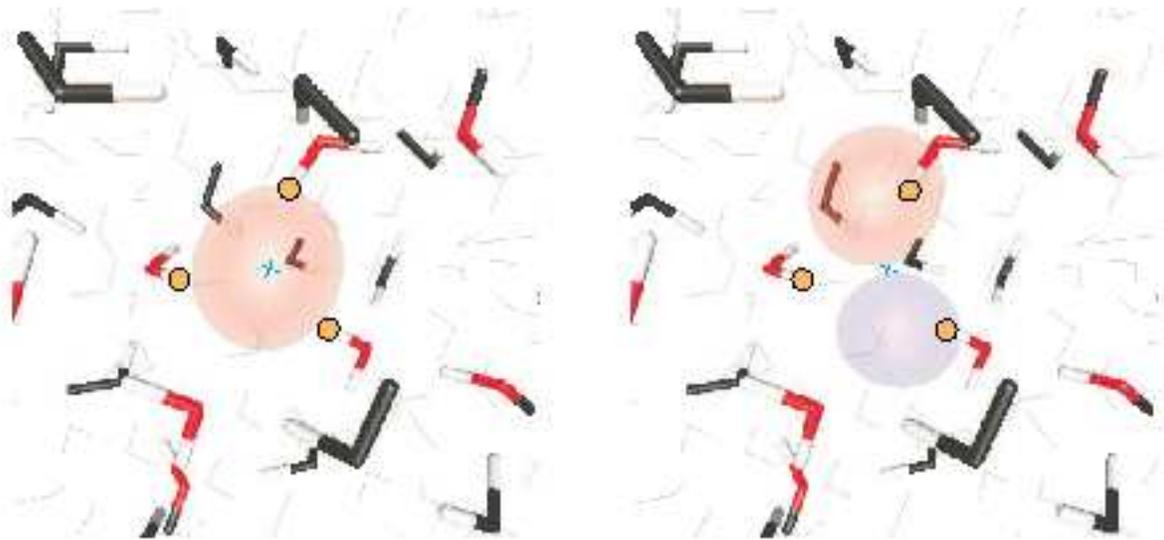

## V⁻, 50 K : FSGO

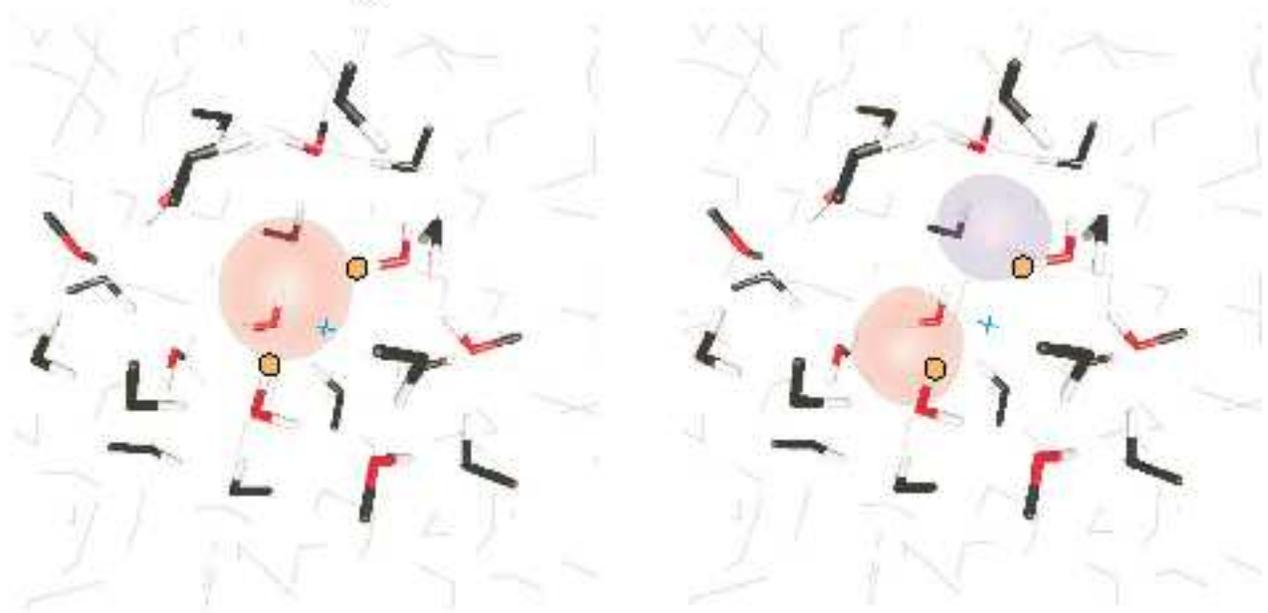

# V⁻, 50 K : FSGO

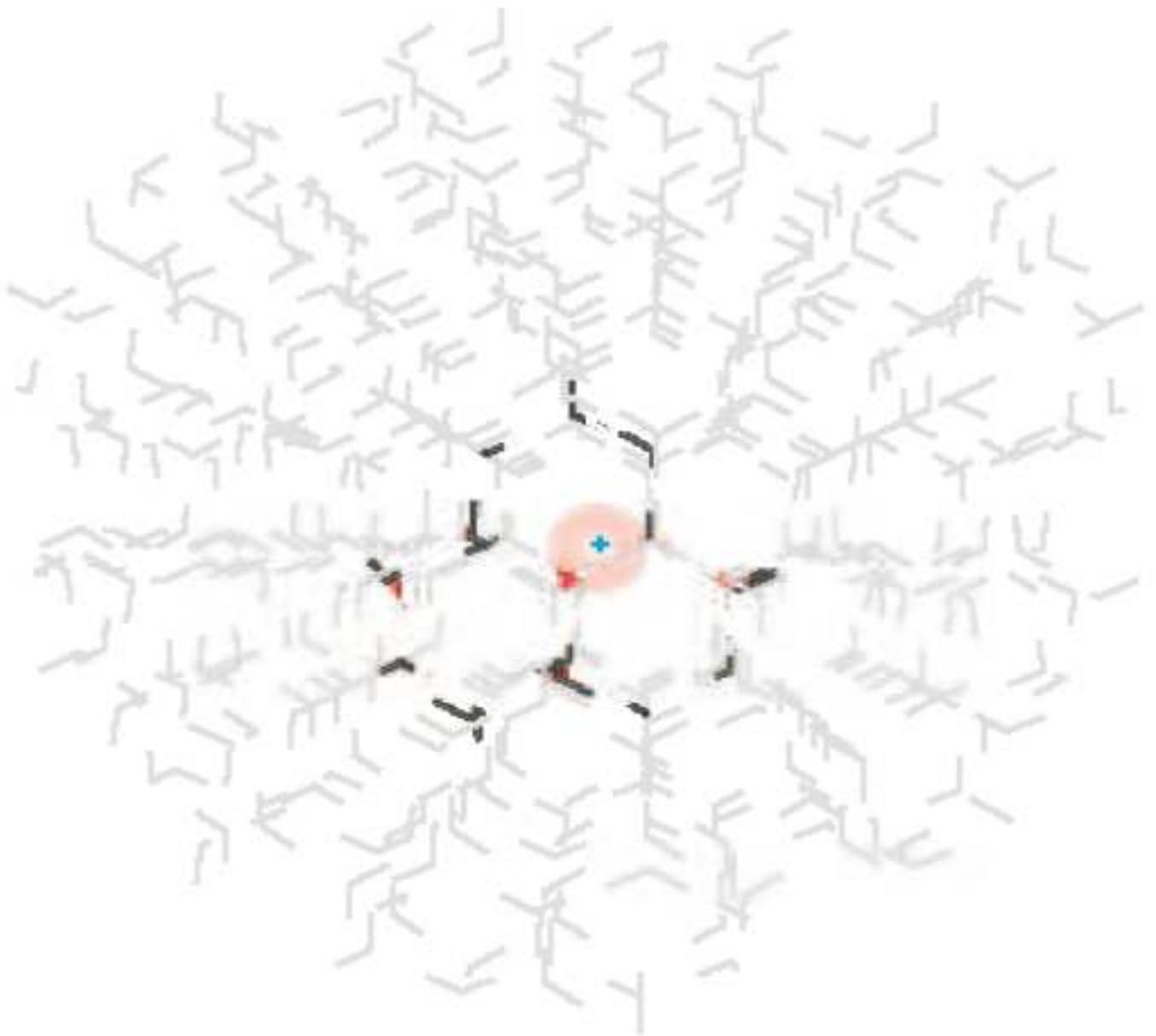

# C⁻, 50 K : FSGO

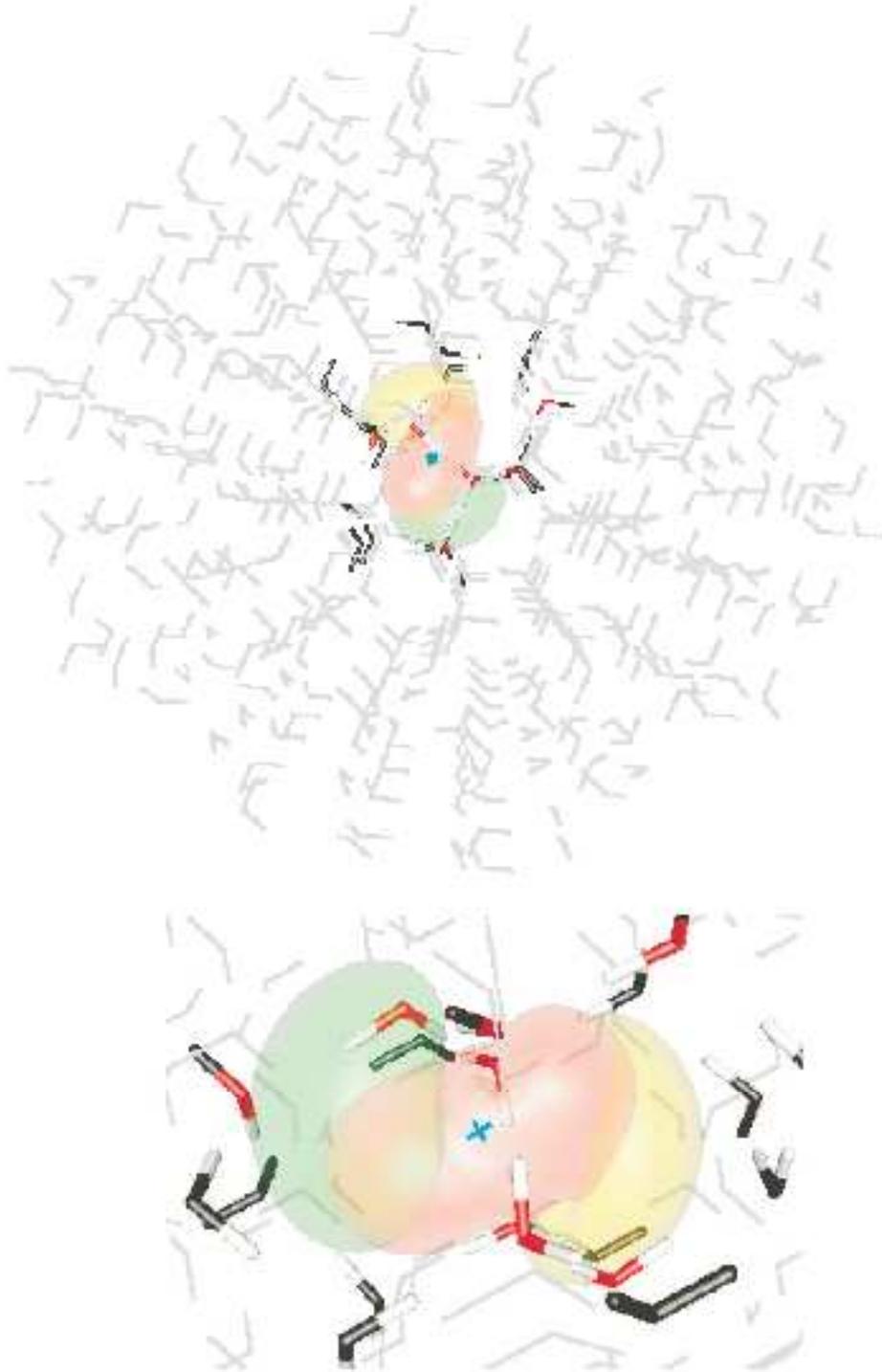

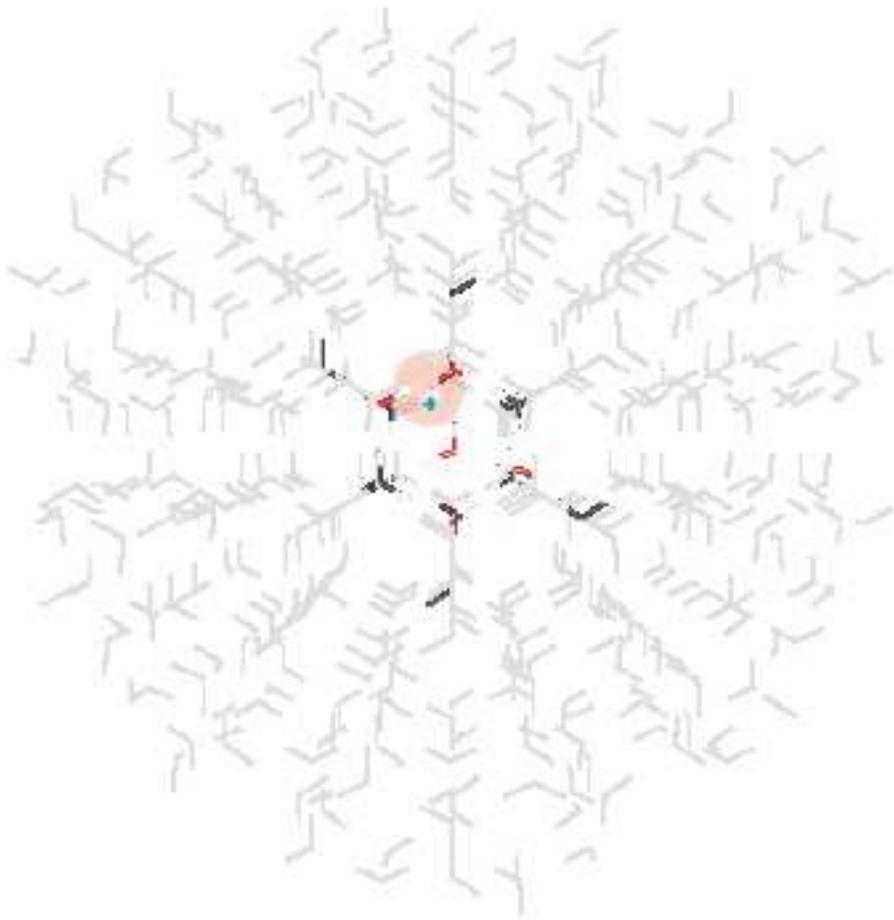

I⁻, 50 K : FSGO

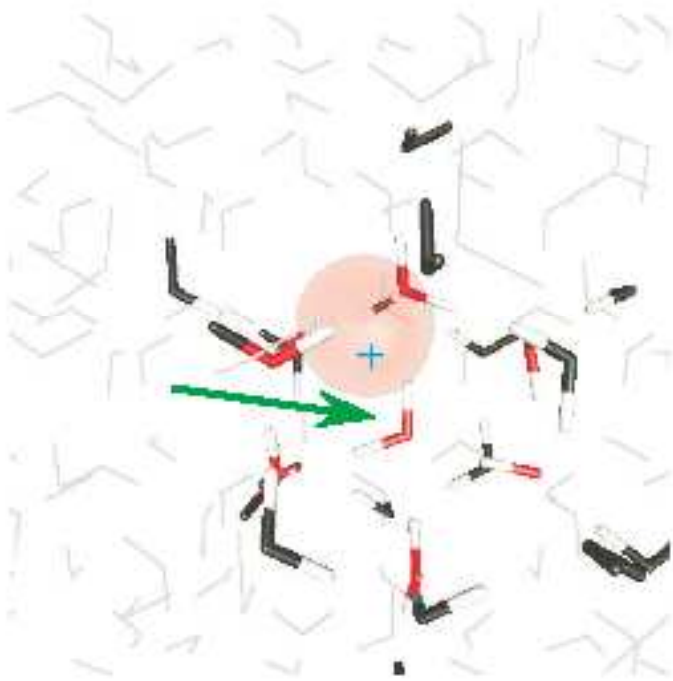

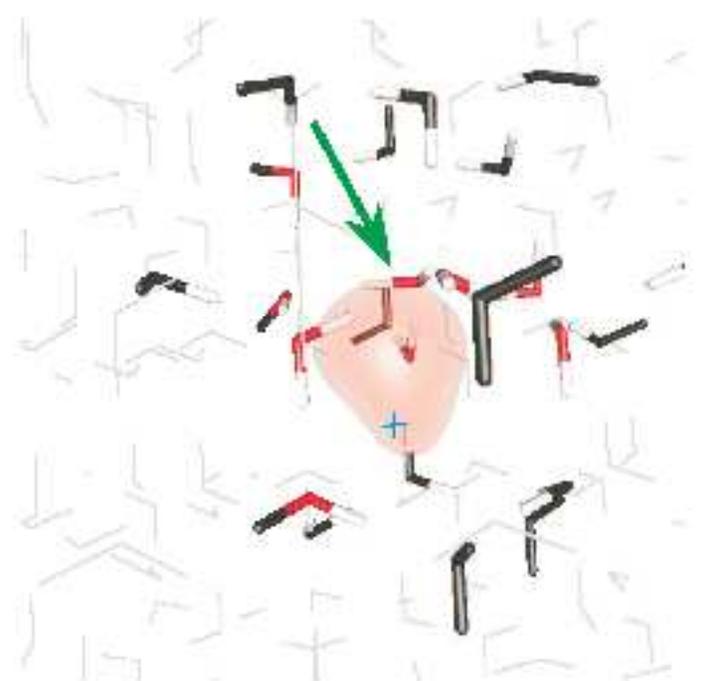

**VD** <sup>-</sup>

**V** <sup>-</sup>

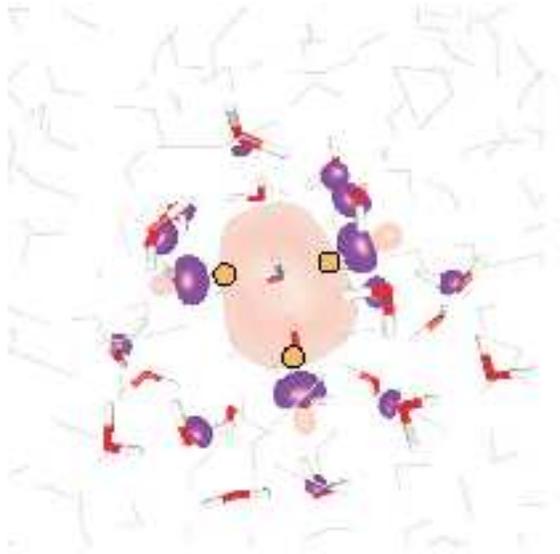 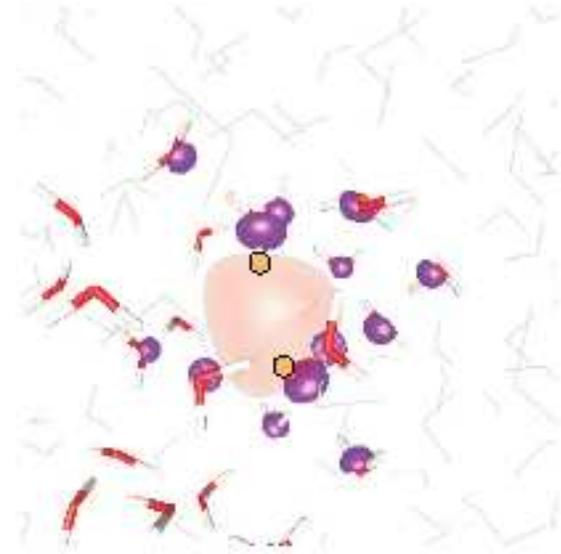

**±0.03**

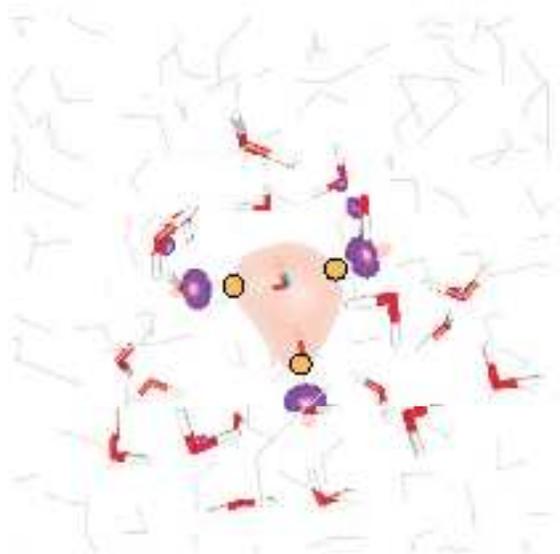 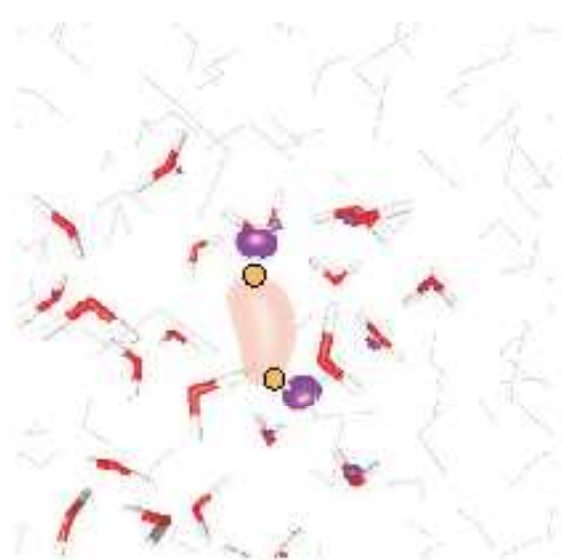

**±0.05**

# VD⁻, 200 K : DFT

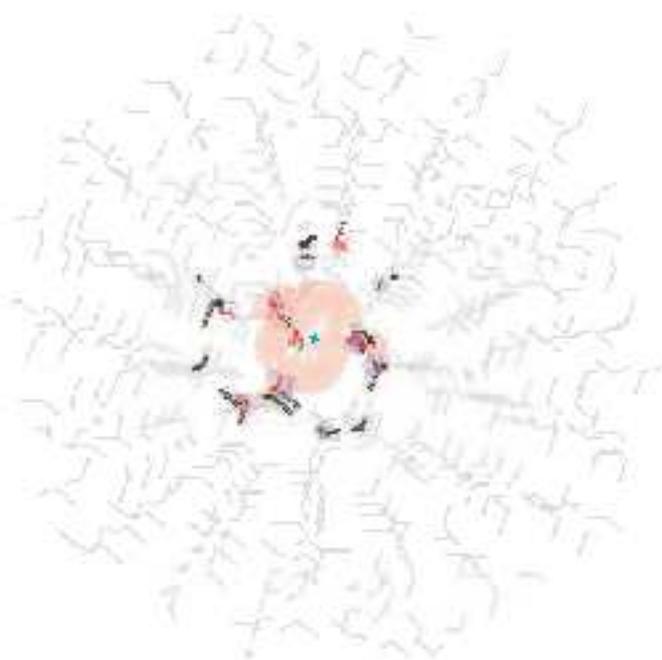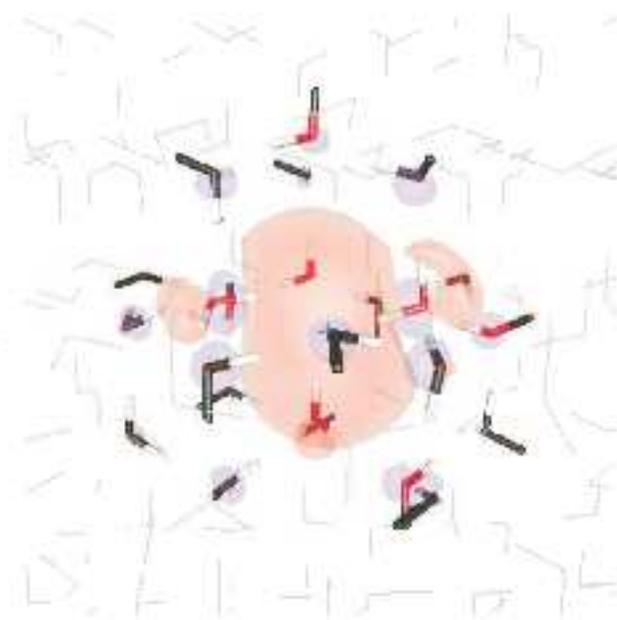

## HOMO ("s")

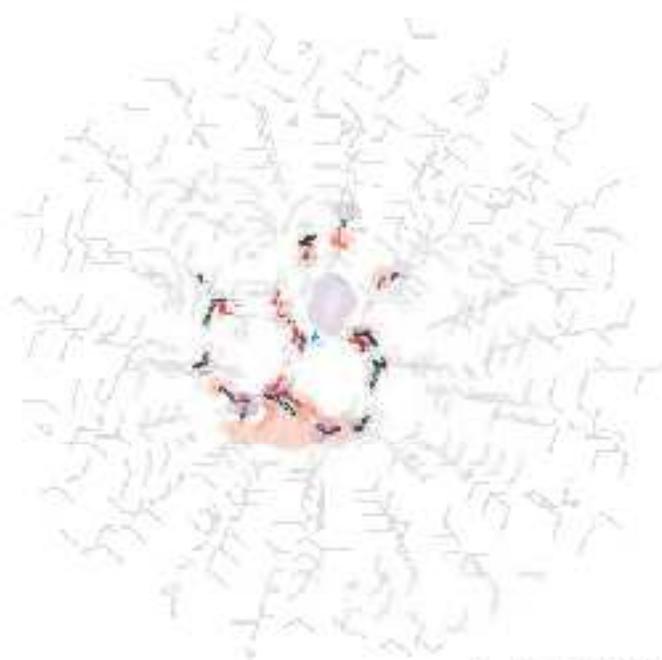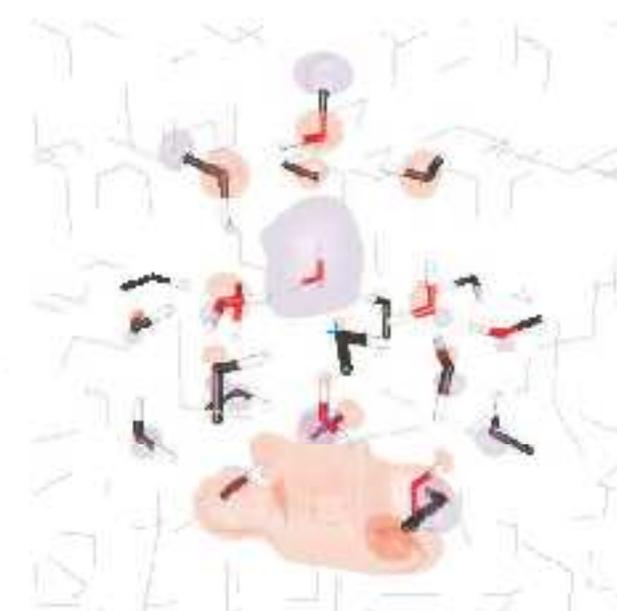

## LUMO ("p")   ±0.02

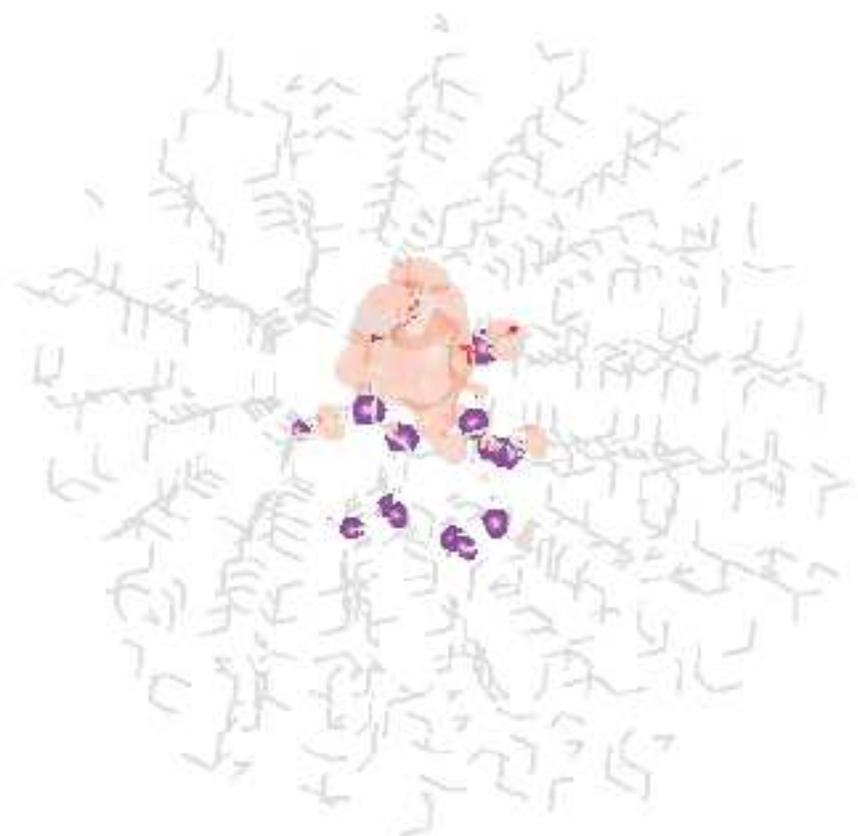

C⁻, 50 K : DFT

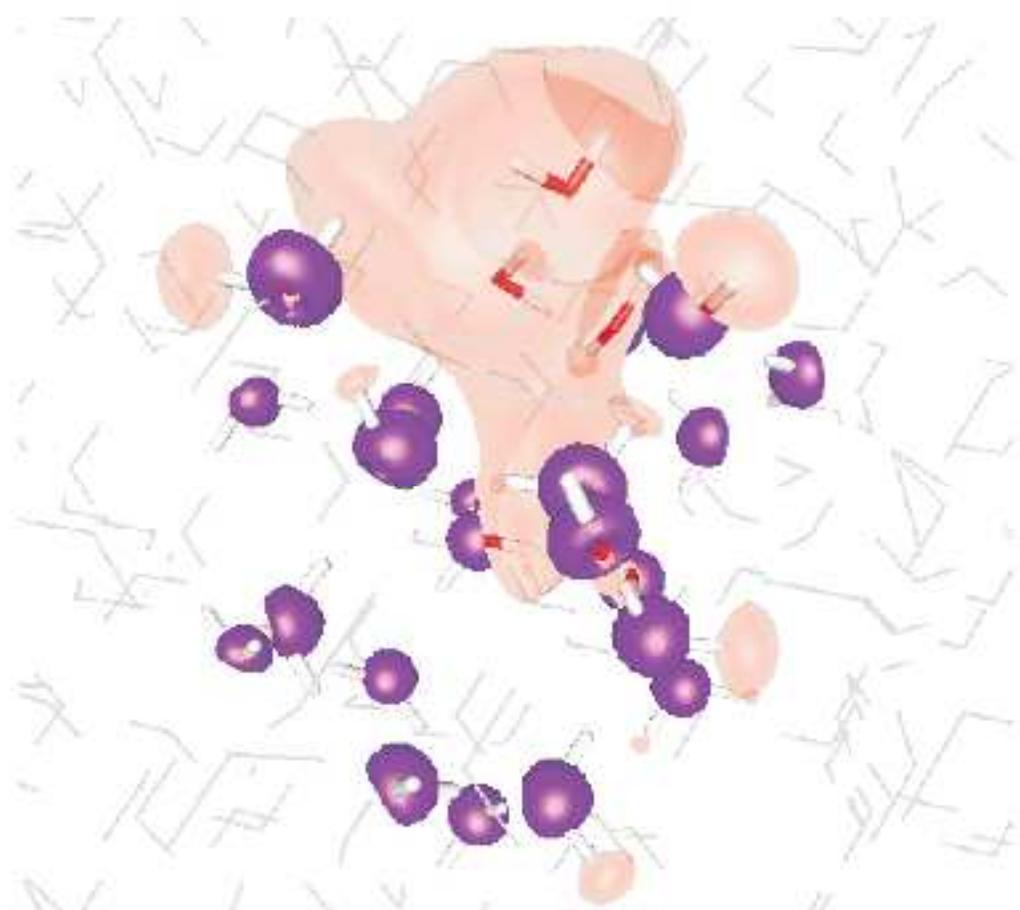

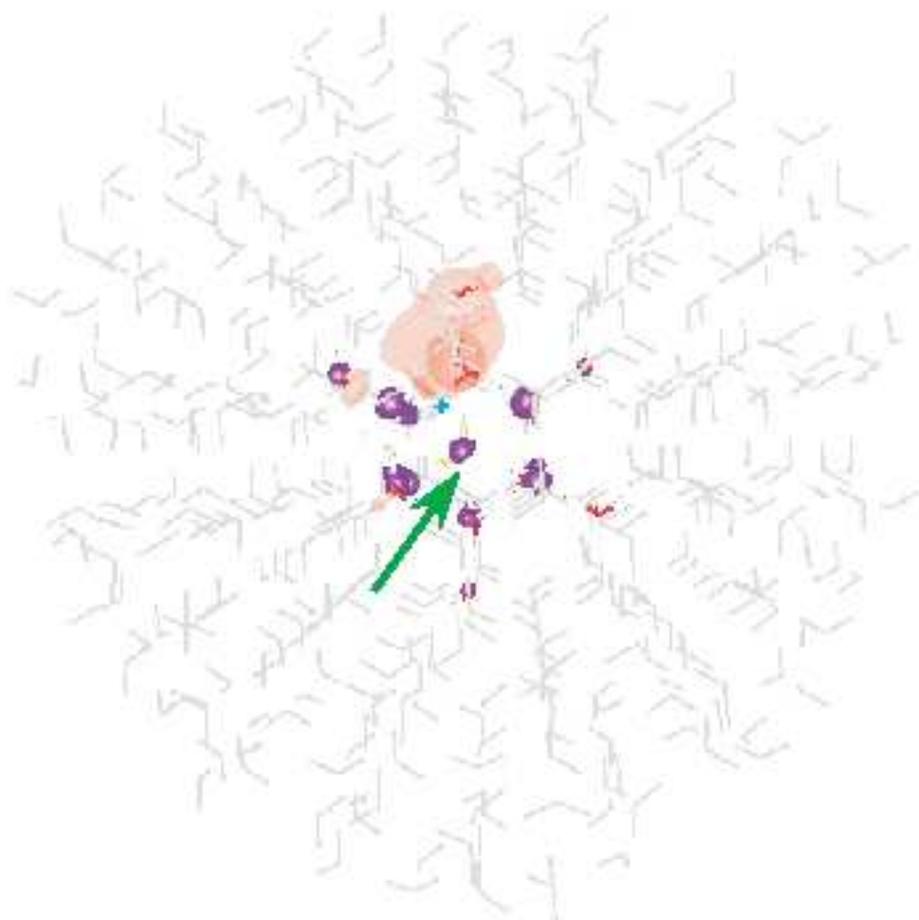

$I^-, 50\,K:$
DFT

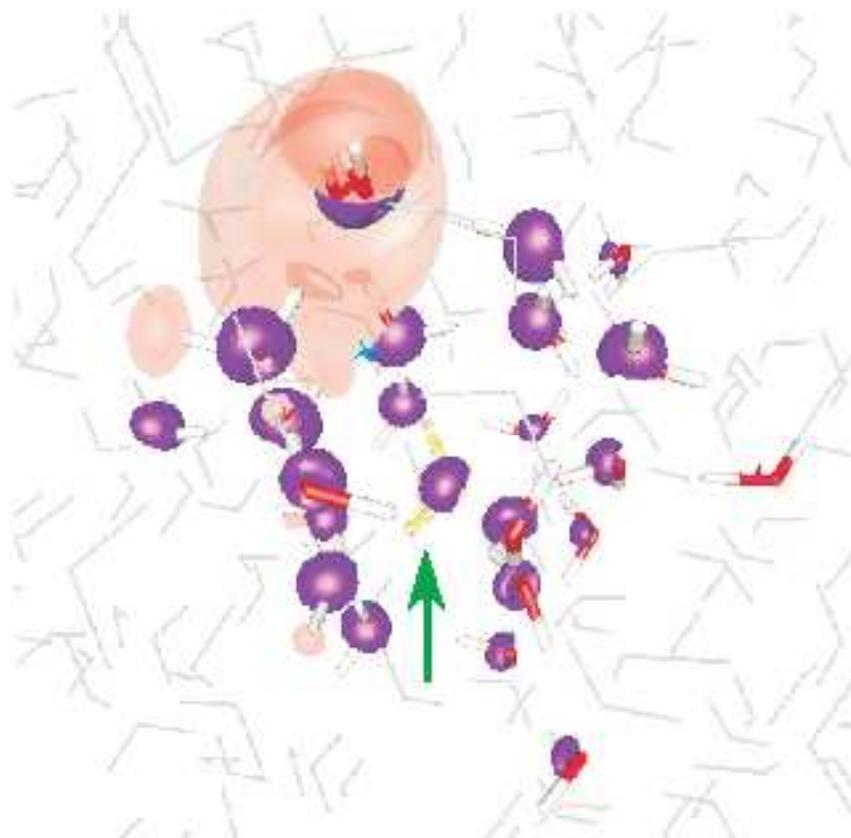

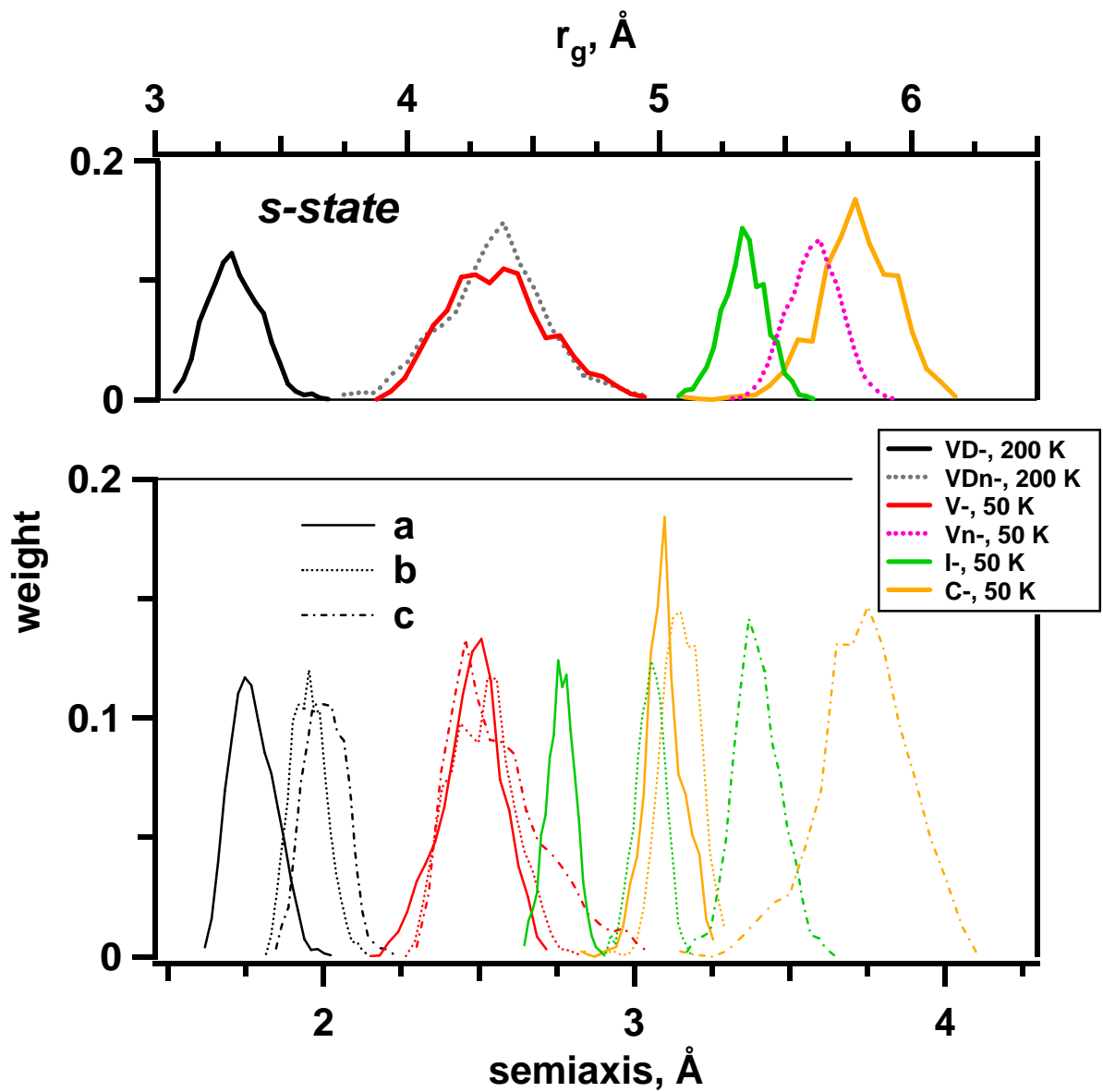

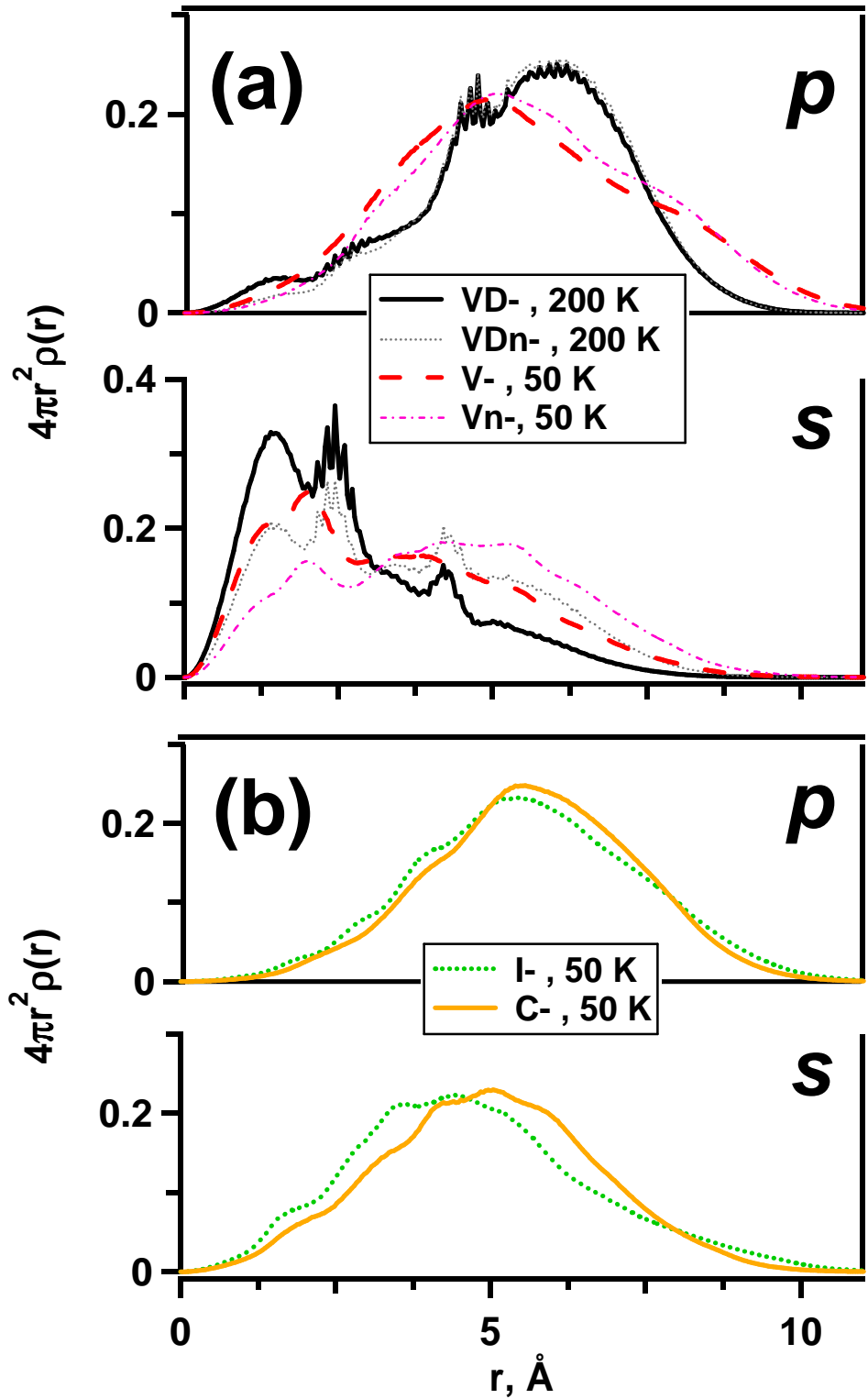

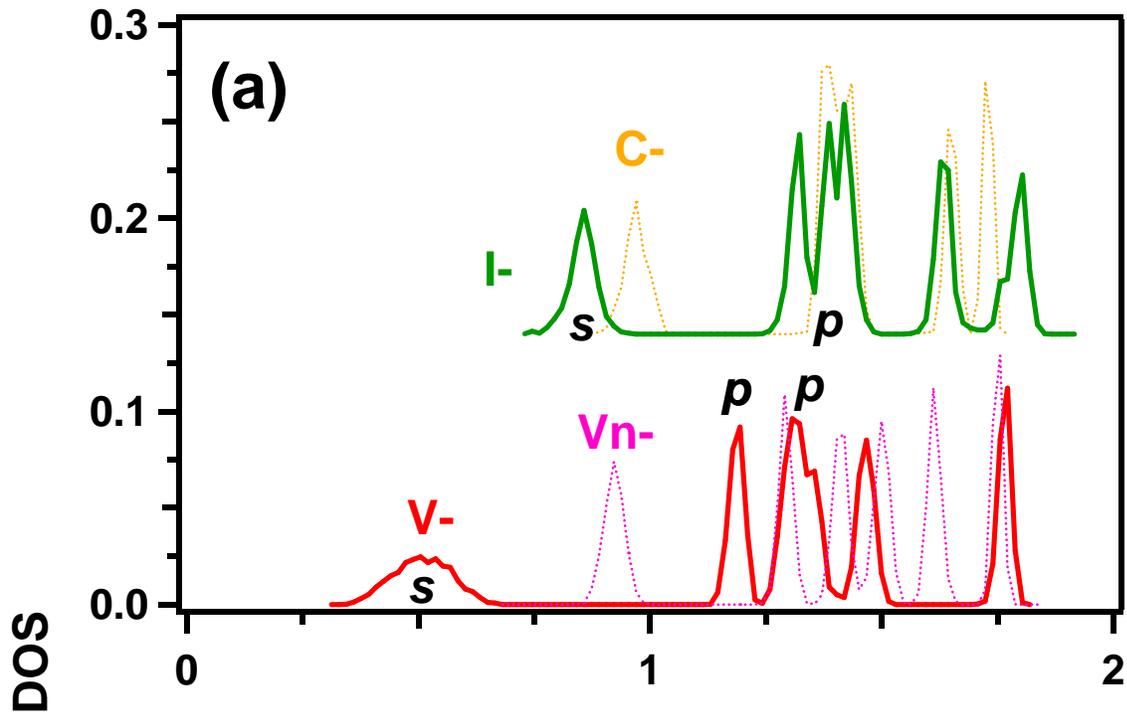

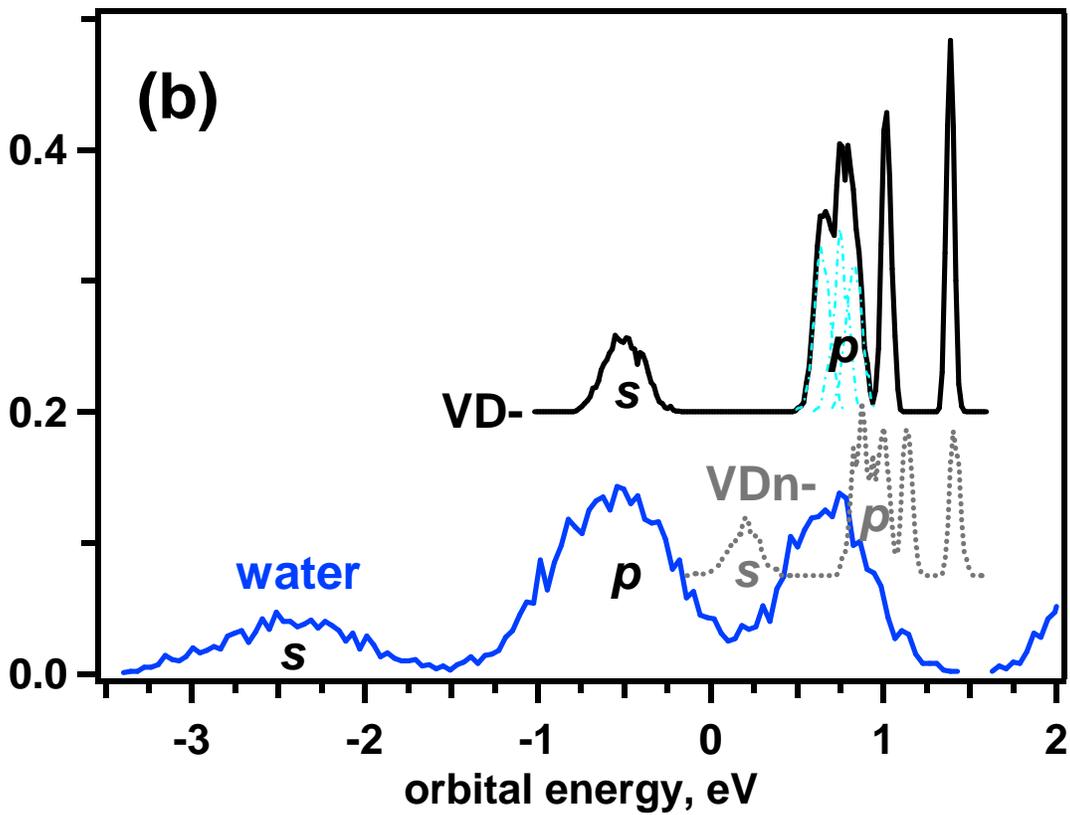

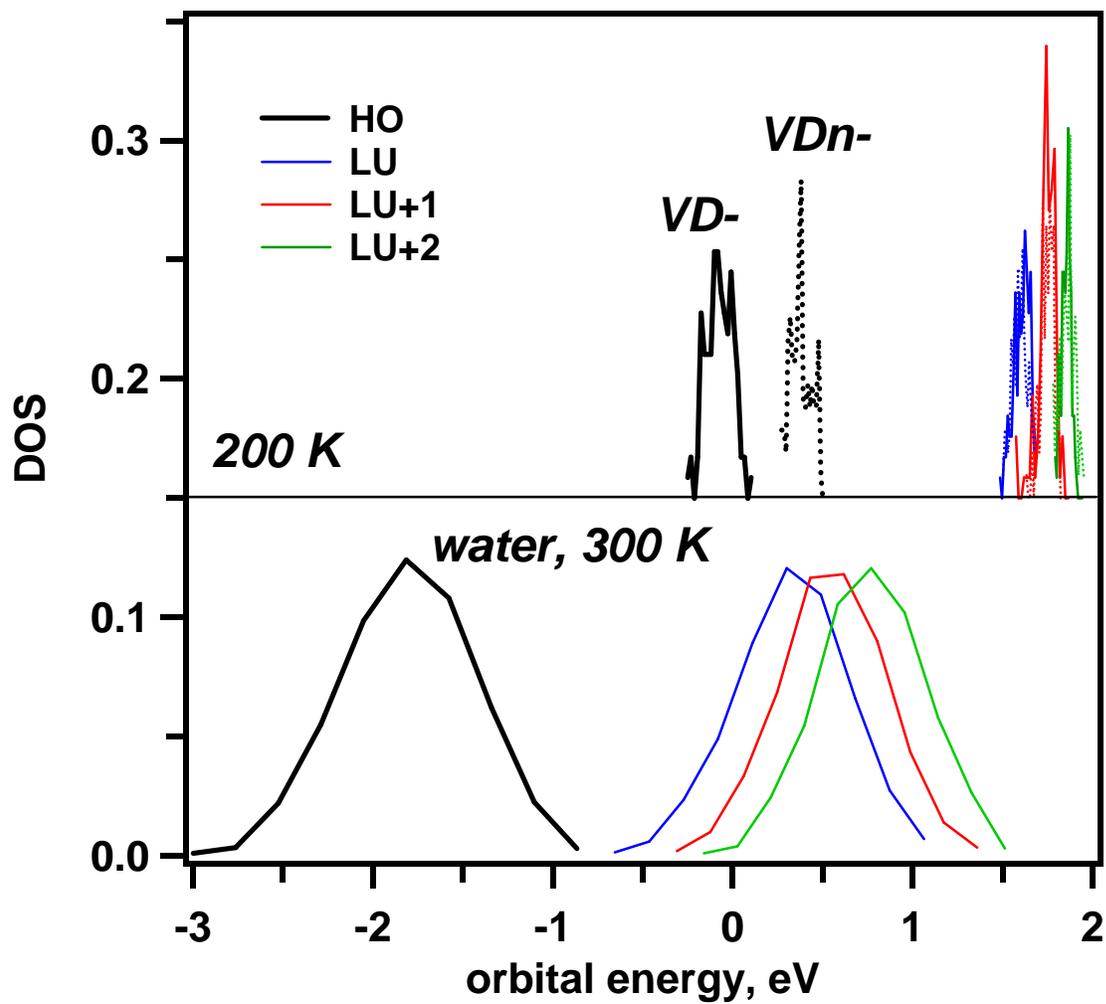

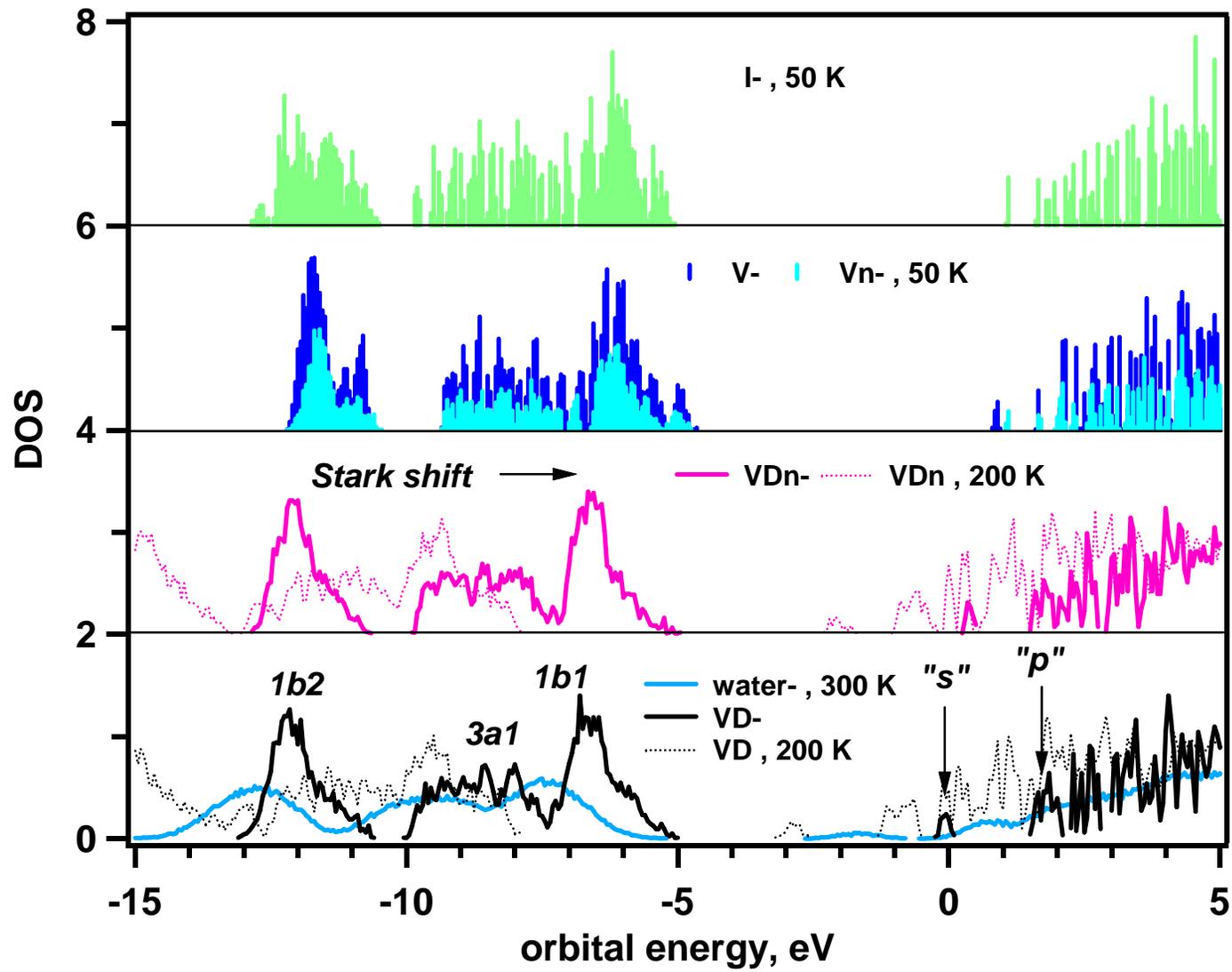